\documentclass[aps,prl,reprint,amsmath,amssymb]{revtex4-1}

\usepackage{bm}
\usepackage[colorlinks=true,citecolor=blue,urlcolor=blue]{hyperref}

\usepackage{pdfpages}

\begin{document}

\title{Correcting inconsistencies in the conventional superfluid path integral scheme}

\author{Brandon M. Anderson}
\author{Rufus Boyack}
\author{Chien-Te Wu}
\author{K. Levin}
\affiliation{James Franck Institute, University of Chicago, Chicago, Illinois 60637, USA}

\begin{abstract}
In this paper we show how to redress a shortcoming of the path integral
scheme for fermionic superfluids and superconductors. This approach
is built around a simultaneous calculation of electrodynamics and
thermodynamics. An important sum rule, the compressibility sum rule,
fails to be satisfied in the usual calculation of the electromagnetic
and thermodynamic response at the Gaussian fluctuation level. Here
we present a path integral scheme to address this inconsistency. Specifically,
at the leading order we argue that the superconducting gap should
be calculated using a different saddle point condition modified by
the presence of an external vector potential. This leads to the well
known gauge-invariant BCS electrodynamic response and is associated
with the usual (mean field) expression for thermodynamics. In this
way the compressibility sum rule is satisfied at the BCS level. Moreover,
this scheme can be readily extended to address arbitrary higher order
fluctuation theories. At any level this approach will lead to a gauge
invariant and compressibility sum rule consistent treatment of electrodynamics
and thermodynamics. 
\end{abstract}

\maketitle

There is a great interest from diverse physics communities in understanding
superfluids~\cite{Giorgini2008,Drummond2009,Hu2012} and superconductors~\cite{Hur2009,Chen2005}
with stronger than BCS correlations. These strong correlations are
present in both high temperature superconductors and in ultra coldourcompanion
Fermi superfluids. At the heart of probes of superfluidity are electrodynamic
and thermodynamic responses. It is, therefore, important to have a
consistent theory for addressing both of these. One consistency requirement
is that of gauge invariance. This affects only the electrodynamics,
and importantly introduces collective modes of the order parameter.
Another consistency requirement involves the inter-connection between
electrodynamics and thermodynamics. This is encapsulated in the compressibility
sum rule~\cite{Pines}.

The path integral scheme is particularly well suited to consistency
checks related to this inter-connection because it \emph{simultaneously}
derives electrodynamics and thermodynamics. However, this scheme,
as it is applied in the literature, is not consistent with the compressibility
sum rule~\cite{ourcompanion}. Indeed, this inconsistency shows up
at the lowest level of approximation needed to arrive at gauge invariant
electrodynamics. Stated more concretely, the path integral approach
raises a natural question: even at the strict BCS level, if fluctuations
around the mean-field solution are necessary for gauge invariant electrodynamics,
should these fluctuations yield additional contributions to thermodynamics
beyond those of the fermionic quasiparticles? Such ``gauge restoring''
terms should have definite observable consequences. For example, in
neutral superfluids (such as superfluid He-3 and atomic Fermi gases)
these phonon modes would lead to power law contributions in measurable
properties such as the specific heat. There seems to be no consensus
about whether these non-BCS terms should or should be not be considered~\cite{Yu2009}.

Here we appeal to the compressibility sum rule to address this question.
We define $\Omega=\Omega_{{\rm mf}}+\Omega_{{\rm fl}}$ as the thermodynamic
potential resulting from a calculation that uses Gaussian fluctuations
(${\rm {fl}}$) around mean field theory (${\rm {mf}}$) to establish
a BCS-level gauge invariant electrodynamic response. We consider $n$
particles having chemical potential $\mu$. Within this formulation,
which we call the gauge restoring Gaussian fluctuation (GRGF) theory,
the number of particles $n=-\partial\Omega/\partial\mu$ has a leading
order mean-field term $n_{{\rm mf}}$ and a fluctuation contribution
$n_{{\rm fl}}$. Similarly the electrodynamic kernel which derives
from $\Omega$ contains the counterpart mean-field and fluctuation
terms, both of which combined lead to a proper gauge invariant BCS
density-density correlation function $K^{00}(\omega,\mathbf{q})$.
One can show that $n=n_{{\rm mf}}+n_{{\rm fl}}$ satisfies 
\begin{eqnarray}
K^{00}(\omega=0,\mathbf{q}\rightarrow0) & = & -\frac{\partial n_{{\rm mf}}}{\partial\mu}\neq-\frac{\partial n}{\partial\mu}.\label{eq:Compress-Sum-Rule}
\end{eqnarray}
This demonstrates an explicit violation~\cite{ourcompanion} of the
compressibility sum rule, which should read $K^{00}(\omega=0,\mathbf{q}\rightarrow0)=-\partial n/\partial\mu$.
It also demonstrates (at least at an empirically suggestive level)
what assumptions need to be made to satisfy the compressibility sum
rule within BCS theory.

In this paper we present a path integral framework modified from that
outlined above. For both the lowest order mean-field, and Gaussian
fluctuation levels, we will derive theories fully consistent with
gauge invariance and the compressibility sum rule. Indeed, this consistency
can in principle be achieved at all orders of approximation within
our path integral re-formulation.

The GRGF approach leading to Eq.~(\ref{eq:Compress-Sum-Rule}) was
presented in a fairly extensive literature~\cite{Altland,Goryo1998a,Lutchyn2008,Ojanen2013,Roy2008,Stone2004,Goryo1998},
where fluctuations of the mean-field phase $\phi$ were used to restore
gauge invariance. These fluctuations enter as a ``dressed'' vector
potential $\tilde{A}_{\mu}=A_{\mu}+\partial_{\mu}\phi$, which is
then expanded to quadratic order. Integration of the fluctuations
$\phi$ resulted in the standard electromagnetic response kernel of
strict BCS theory. We emphasize here~\cite{Altland,Goryo1998a,Lutchyn2008,Ojanen2013,Roy2008,Stone2004,Goryo1998}
that the focus was on electrodynamics while the thermodynamic implications
were of no concern.

In contrast, understanding thermodynamics associated with Gaussian
fluctuation theories (beyond the BCS level) was the focus of work
by a different community, that studying ultracold Fermi superfluids~\cite{Diener2008,He2015a,Loktev2000,Ohashi2003,Perali2004,Pieri2000,Pieri2005}.
In these neutral superfluids, soft bosonic collective modes arising
from fluctuations were shown to provide new thermodynamic contributions
in addition to those of the fermionic quasi-particles of BCS theory.

Yet another series of studies incorporated these Gaussian-level (beyond
BCS) fluctuations to revisit electrodynamics in a higher level theory.
By introducing a small phase twist in the thermodynamic potential,
it was argued that one could determine the superfluid density $\rho_{s}$~\cite{Taylor2006,Fukushima2007,Taylor2010};
moreover, this now contained bosonic contributions, not present in
BCS theory. These were somewhat similar (but not equivalent) to contributions
found~\cite{Perali2004,Pieri2000} within a very different diagrammatic
formalism. 

All this previous literature relating to Gaussian fluctuations can
be summarized by noting that there have been separate path integral
studies of superfluid electrodynamics and of thermodynamics. What
is missing is an analysis of the constraints which relate the two.
In this paper we address this shortcoming.

\paragraph*{Path integral and mean field.--}

Here we consider a fermionic partition function for a neutral, attractive,
Fermi gas with $s$-wave pairing. The techniques presented here can
be readily extended to higher order pairing, and Coulomb interactions
can be included at the RPA level~\cite{Lutchyn2008}. The partition
function is calculated using the Hubbard-Stratonovich (HS) path integral
\begin{equation}
\mathcal{Z}\left[A\right]=\int\mathcal{D}\left[\bm{\Delta}\right]e^{-S_{{\rm HS}}\left[\bm{\Delta},A\right]},\label{eq:Za}
\end{equation}
where the HS action takes the usual form $S_{{\rm HS}}\left[\bm{\Delta},A\right]=\int dx\frac{\left|\Delta\right|^{2}}{g}-{\rm Tr}\ln\left[-\mathcal{G}^{-1}\left[\bm{\Delta},A\right]\right]$~\cite{Altland,Fradkin},
$g>0$ is an interaction constant, and ${\rm Tr}\left[\cdot\right]$
includes a trace over both position and Nambu indices; throughout
we set $\hbar=k_{B}=1$. The inverse Nambu Green's function $\mathcal{G}^{-1}\left[\bm{\Delta},A\right]=\mathcal{G}_{0}^{-1}\left[A\right]-\Sigma\left[\bm{\Delta}\right]$
is constructed from a single particle Green's function $\mathcal{G}_{0}\left[A\right]$
and a self-energy $\Sigma=-\bm{\Delta}\cdot\bm{\tau}$, with $\bm{\tau}=\left(\tau_{1},\tau_{2}\right)$
a vector of Nambu Pauli matrices. Throughout we use the notation $\bm{\Delta}=\left(\Delta_{1},\Delta_{2}\right)$
to represent two real HS fields $\Delta_{a}\left(x\right)$, with
$a=1,2$, consistent with previous literature~\footnote{This notation is equivalent to that in Refs.~\cite{Kulik1981,Kosztin2000,Guo2013,ourcompanion}.
The reality condition is expressed in position space; a momentum-space
parametrization has a different condition~\cite{Guo2013}. The conventional
BCS self energy $\Sigma\left[\bm{\Delta}\right]=-\left(\Delta\tau_{+}+\Delta^{*}\tau_{-}\right)$
suggests an equivalent complex parametrization $\Delta_{\pm}=\Delta_{1}\pm i\Delta_{2}$
where the BCS gap is identified through $\Delta\equiv\Delta_{-}$.
See the Supplemental Material~\cite{supplement} for details.}. The single particle Green's function $\mathcal{G}_{0}\left[A\right]$
is kept general, but we note that an electromagnetic vector potential
$A_{\mu}$ has been explicitly included at this level.

We now calculate $\mathcal{Z}\left[A\right]$ at the mean-field level
using the saddle point approximation $\delta S_{{\rm HS}}\left[\bm{\Delta},A\right]/\delta\Delta_{a}=0$
in the presence of $A_{\mu}\neq0$. This is to be contrasted with
previous work (belonging to the GRGF scheme)~\cite{Altland,Goryo1998a,Lutchyn2008,Ojanen2013,Roy2008,Stone2004,Goryo1998}
where the saddle point condition assumed $A_{\mu}=0$. Here, explicit
calculation produces the standard BCS gap equation, $0=2\Delta_{a}\left[A\right]/g-{\rm Tr}\left[\mathcal{G}\left[\bm{\Delta}\left[A\right],A\right]\tau_{a}\right]$,
\emph{in the presence of a non-zero vector potential $A_{\mu}$.}
We define the solution to this gap equation as $\bm{\Delta}^{{\rm mf}}\left[A\right]$,
which depends on $A_{\mu}$. We note that a different community has
exploited the advantages of considering alternative saddle point schemes~\cite{Kamenev1999}.

At the present mean-field (saddle point) level, we can write $\mathcal{Z}_{{\rm mf}}\left[\bm{\Delta}^{{\rm mf}}\left[A\right],A\right]=e^{-S_{{\rm mf}}}$,
where the mean-field action $S_{{\rm mf}}=S_{{\rm HS}}\left[\bm{\Delta}^{{\rm mf}}\left[A\right],A\right]$
is the HS action evaluated at the solution to the saddle point equations.
In general we cannot explicitly calculate the solution to the gap
equation for $A_{\mu}\neq0$. Instead, we will first use the self-consistent
gap equation to find the variation of $\bm{\Delta}^{{\rm mf}}\left[A\right]$
with respect to a variation in $A_{\mu}$. We then take the $A_{\mu}\rightarrow0$
limit, after which all quantities are calculated using $\bm{\Delta}^{{\rm mf}}\equiv\bm{\Delta}^{{\rm mf}}[0]$.
Thus, no additional computational difficulties arise when using this
self-consistency condition compared to the GRGF formalism.

\paragraph*{Response functions at saddle point level.--}

Given an arbitrary ``effective action'' $S_{{\rm eff}}\left[A\right]=-\ln\mathcal{Z}\left[A\right]$
in the presence of a weak perturbation $A_{\mu}$, the response kernel
comes from the second functional derivative of the action in the $A_{\mu}\rightarrow0$
limit~\cite{Fradkin}. As such, we can expand $S_{{\rm eff}}\left[A\right]\approx S_{{\rm eff}}\left[0\right]+\frac{1}{2}\int dx\int dx^{\prime}A_{\mu}\left(x\right)K^{\mu\nu}\left(x,x^{\prime}\right)A_{\nu}\left(x^{\prime}\right)$
to second order in the vector potential $A_{\mu}$, where 
\begin{equation}
K^{\mu\nu}\left(x,x^{\prime}\right)=\left.\frac{\delta^{2}S_{{\rm eff}}\left[A\right]}{\delta A_{\mu}\left(x\right)\delta A_{\nu}\left(x^{\prime}\right)}\right|_{A\rightarrow0}\label{eq:KAA}
\end{equation}
is the response kernel for an arbitrary action $S_{{\rm eff}}\left[A\right]$.

We now calculate the mean-field response using the definition in Eq.~(\ref{eq:KAA})
by including a nonzero vector potential in the saddle point condition,
i.e., replace $S_{{\rm eff}}\left[A\right]$ by $S_{{\rm mf}}=S_{{\rm mf}}\left[\bm{\Delta}^{{\rm mf}}\left[A\right],A\right]$.
When taking a functional derivative with respect to $A_{\mu}$, new
terms arise from a ``functional chain rule''~\cite{Altland} applied
to the self-consistent gap $\bm{\Delta}^{{\rm mf}}\left[A\right]$.
These terms, which do not not emerge for a gap calculated around $A_{\mu}=0$
as in GRGF, are crucial for maintaining gauge invariance. The full
response kernel then takes the form: 
\begin{eqnarray}
K_{{\rm mf}}^{\mu\nu}(x,x^{\prime}) & = & \left.\frac{\delta^{2}S_{{\rm mf}}}{\delta A_{\mu}^{x}\delta A_{\nu}^{x^{\prime}}}\right|_{\bm{\Delta}^{{\rm mf}}}+\frac{\delta\Delta_{a}^{y}}{\delta A_{\mu}^{x}}\left.\frac{\delta^{2}S_{{\rm mf}}}{\delta\Delta_{a}^{y}\delta\Delta_{b}^{y^{\prime}}}\right|_{\bm{\Delta}^{{\rm mf}}}\frac{\delta\Delta_{b}^{y^{\prime}}}{\delta A_{\nu}^{x^{\prime}}}\nonumber \\
 & + & \frac{\delta\Delta_{a}^{y}}{\delta A_{\mu}^{x}}\left.\frac{\delta^{2}S_{{\rm mf}}}{\delta\Delta_{a}^{y}\delta A_{\nu}^{x^{\prime}}}\right|_{\bm{\Delta}^{{\rm mf}}}+\left.\frac{\delta^{2}S_{{\rm mf}}}{\delta A_{\mu}^{x}\delta\Delta_{a}^{y}}\right|_{\bm{\Delta}^{{\rm mf}}}\frac{\delta\Delta_{a}^{y}}{\delta A_{\nu}^{x^{\prime}}}\nonumber \\
 & + & \left.\frac{\delta S_{{\rm mf}}}{\delta\Delta_{a}^{y}}\right|_{\bm{\Delta}^{{\rm mf}}}\frac{\delta^{2}\Delta_{a}^{y}}{\delta A_{\mu}^{x}\delta A_{\nu}^{x^{\prime}}},\label{eq:Ksp-general}
\end{eqnarray}
where the $A_{\mu}\rightarrow0$ limit is applied after taking all
derivatives. In this equation we have introduced the notation $\Delta_{a}^{x}\equiv\Delta_{a}\left(x\right)$
and $A_{\mu}^{x}\equiv A_{\mu}\left(x\right)$; repeated subscript
(superscript) indices $a,b$ ($y,y^{\prime}$) should be interpreted
as an implied Einstein summation (integration.)

To express Eq.~(\ref{eq:Ksp-general}) in a more suggestive form,
we define the set of two-point response functions~\cite{Kulik1981,Kosztin2000,Guo2013,ourcompanion}:
\begin{eqnarray}
\mathcal{Q}_{{\rm mf}}^{\alpha\beta}\left(x,x^{\prime}\right) & \equiv & \left.\frac{\delta^{2}S_{{\rm mf}}\left[\bm{\Delta}^{{\rm mf}},A\right]}{\delta\mathcal{A}_{\alpha}\left(x\right)\delta\mathcal{A}_{\beta}\left(x^{\prime}\right)}\right|_{A\rightarrow0},\label{eq:Resp1-1}
\end{eqnarray}
where $\mathcal{A}_{\alpha}=\left(\Delta_{1}^{{\rm mf}},\Delta_{2}^{{\rm mf}},A_{\mu}\right)$
parameterizes both gap and vector potential response. The kernel $K_{0,{\rm mf}}^{\mu\nu}\equiv\mathcal{Q}_{{\rm mf}}^{\mu\nu}$
is the standard (non-gauge invariant) response as calculated with
a gap $\bm{\Delta}^{{\rm mf}}$; the functions $Q_{{\rm mf}}^{a\mu}=\mathcal{Q}_{{\rm mf}}^{a\mu}$
and $Q_{{\rm mf}}^{ab}=\mathcal{Q}_{{\rm mf}}^{ab}$ come from ``partial''
derivatives in the functional chain rule. We note that the propagator
$Q_{{\rm mf}}^{ab}$ is equivalent to a ``$GG$'' $t$-matrix theory
for a BCS self-energy, and therefore can be interpreted as an emergent
bosonic propagator~\cite{Chen2005,Diener2008}. Using these definitions,
the mean-field level gauge invariant response is compactly written
\begin{equation}
K_{{\rm mf}}^{\mu\nu}=K_{0,{\rm mf}}^{\mu\nu}+\Pi_{a}^{\mu}Q_{{\rm mf}}^{a\nu}+Q_{{\rm mf}}^{\mu a}\Pi_{a}^{\nu}+\Pi_{a}^{\mu}Q_{{\rm mf}}^{ab}\Pi_{b}^{\nu},\label{eq:Ksp-schematic}
\end{equation}
where we henceforth include an implicit integration over $y,y^{\prime}$
for every Einstein summation over $a,b$. In Eq.~(\ref{eq:Ksp-schematic})
we have introduced the collective mode terms $\Pi_{a}^{\mu}\left(x,x^{\prime}\right)\equiv\delta\Delta_{a}^{{\rm mf}}\left[A\right]\left(x^{\prime}\right)/\delta A_{\mu}\left(x\right)$;
these explicitly restore gauge invariance beyond the ``bubble''
response kernel $K_{0,{\rm mf}}^{\mu\nu}$~\cite{Kulik1981,Kosztin2000,Guo2013,ourcompanion}.
In the saddle point response, the third line in Eq.~(\ref{eq:Ksp-general})
vanishes.

Using the revised saddle point condition, along with the above definitions,
the collective modes are $\Pi_{a}^{\mu}=-\left[Q_{{\rm mf}}^{ab}\right]^{-1}Q_{{\rm mf}}^{b\mu}$
where the inverse $\left[Q_{{\rm mf}}^{ab}\right]^{-1}$ is taken
over both position and Nambu indices (see Supplemental Material~\cite{supplement}).
We emphasize that these collective modes are associated with the mean-field
level of approximation. Finally, after taking the $A_{\mu}\rightarrow0$
limit, the momentum space response is 
\begin{equation}
K_{{\rm mf}}^{\mu\nu}\left(q\right)=K_{0,{\rm mf}}^{\mu\nu}\left(q\right)-Q_{{\rm mf}}^{\mu a}\left(-q\right)\left[Q_{{\rm mf}}^{ab}\left(q\right)\right]^{-1}Q_{{\rm mf}}^{b\nu}\left(q\right).\label{eq:Ksp-only}
\end{equation}
This is the usual gauge invariant response kernel in BCS theory~\cite{Kulik1981}
which includes both amplitude and phase collective modes.

Importantly, the response kernel $K_{{\rm mf}}^{\mu\nu}$, which is
explicitly gauge invariant, was obtained without including Gaussian
fluctuations, which are usually invoked in the GRGF literature. In
this way \emph{ }\textit{the self-consistent treatment of the gap
in the presence of a vector potential restores gauge invariance at
the mean-field level.} Because there are no accompanying bosonic degrees
of freedom in the thermodynamics, the compressibility sum rule will
be shown to be exactly satisfied using this method, in contrast to
the more conventional path integral methodology.

\paragraph*{Beyond saddle point.--}

Often it is desirable to calculate the path integral beyond the saddle
point approximation. In order to do this, one changes variables from
the HS field $\bm{\Delta}$ to a fluctuation $\bm{\eta}=\left(\eta_{1},\eta_{2}\right)$
around the saddle point solution defined through $\bm{\Delta}=\bm{\Delta}^{{\rm mf}}\left[A\right]+\bm{\eta}$.
We note that since $\bm{\eta}$ is a dynamical variable it does not
have any dependence on $A_{\mu}$. The full action is then expressed
exactly as $S_{{\rm HS}}\left[\bm{\Delta},A\right]=S_{{\rm mf}}+S_{\eta}$,
where the action $S_{\eta}\equiv S_{\eta}\left[\bm{\Delta}^{{\rm mf}}\left[A\right],A,\bm{\eta}\right]=S_{{\rm HS}}\left[\bm{\Delta}^{{\rm mf}}\left[A\right]+\bm{\eta},A\right]-S_{{\rm HS}}\left[\bm{\Delta}^{{\rm mf}}\left[A\right],A\right]$
is $\mathcal{O}\left(\eta^{2}\right)$ or higher, since any term linear
in $\bm{\eta}$ vanishes by the saddle point condition. This definition
allows for the exact factorization of the partition function $\mathcal{Z}\left[A\right]=\mathcal{Z}_{{\rm mf}}\left[\bm{\Delta}^{{\rm mf}}\left[A\right],A\right]\mathcal{Z}_{{\rm fl}}\left[\bm{\Delta}^{{\rm mf}}\left[A\right],A\right]$,
where 
\begin{equation}
\mathcal{Z}_{{\rm fl}}\left[\bm{\Delta}^{{\rm mf}}\left[A\right],A\right]=\int\mathcal{D}\left[\bm{\eta}\right]e^{-S_{\eta}\left[\bm{\Delta}^{{\rm mf}}\left[A\right],A,\bm{\eta}\right]}
\end{equation}
is the contribution due to fluctuations beyond mean field.

In calculations of response beyond saddle point, one uses Eq.~(\ref{eq:KAA})
with an effective action $S_{{\rm eff}}\left[A\right]=-\ln\mathcal{Z}\left[A\right]=S_{{\rm mf}}+S_{{\rm fl}}$,
and the fluctuation action $S_{{\rm fl}}=-\ln\mathcal{Z}_{{\rm fl}}\left[\bm{\Delta}^{{\rm mf}}\left[A\right],A\right]$
also depends on the self-consistent gap $\bm{\Delta}^{{\rm mf}}\left[A\right]$.
The response kernel is linear in the action, so that $K^{\mu\nu}=K_{{\rm mf}}^{\mu\nu}+K_{{\rm fl}}^{\mu\nu}$,
where the mean-field response is given in Eq.~(\ref{eq:Ksp-only}).
The new contribution to the response, $K_{{\rm fl}}^{\mu\nu}$, has
a form identical to Eq.~(\ref{eq:Ksp-general}), only with $S_{{\rm mf}}$
replaced by $S_{{\rm fl}}$. Note, however, that the collective mode
terms $\Pi_{a}^{\mu}$ still arise from the mean field self-consistent
gap condition; these collective modes are always constructed from
the $Q_{{\rm mf}}$ propagators, and not from an analogous $Q_{{\rm fl}}$.

This higher order fluctuation response again contains a ``bubble''
term $K_{0,{\rm fl}}^{\mu\nu}$ that arises from bosonic fluctuations.
On its own, $K_{0,{\rm fl}}^{\mu\nu}$ is not gauge invariant. Analogous
to the saddle-point response, the collective modes $\Pi_{\mu}^{a}$,
along with the corresponding $Q_{{\rm fl}}$ response functions, are
necessary to restore gauge invariance. To show that this arbitrary
fluctuation theory is fully gauge invariant, one can verify that $\partial_{\mu}K_{{\rm fl}}^{\mu\nu}=0$
is satisfied (see the Supplemental Material~\cite{supplement}.)
In this way, gauge invariance holds term by term in the expansion
of the action beyond mean-field. This calculation scheme for gauge
invariant response beyond-BCS is a completely general sum rule consistent
scheme and a central result of this manuscript.

\paragraph*{Compressibility sum rule.--}

Thermodynamic quantities can be calculated from derivatives of the
thermodynamic potential, $\Omega=-T\ln\mathcal{Z}=TS_{{\rm eff}}$,
which is the effective action up to the prefactor $T$. Since electromagnetic
response functions also come from derivatives of the effective action,
it is clear that there should be an intimate connection between the
two. An important requirement for consistency between electrodynamics
and thermodynamics is contained in the compressibility sum rule: $\partial n/\partial\mu=-K^{00}\left(0,\mathbf{q}\rightarrow0\right)$.

A formal derivation of this sum rule, for the exact action, arises
from twice invoking the identity $\int dx\,\delta\mathcal{G}_{0}^{-1}/\delta A_{0}\left(x\right)=-\partial\mathcal{G}_{0}^{-1}/\partial\mu$
on the partition function in Eq.~\eqref{eq:Za}. A more intuitive
derivation of this sum rule follows from the fermionic path integral,
before applying the HS transformation. The atom number is $n\equiv\left\langle \int dx\,\hat{n}\left(x\right)\right\rangle =-\partial\Omega/\partial\mu$,
where $\hat{n}\left(x\right)=\sum_{s=\uparrow,\downarrow}\psi_{s}^{\dagger}\left(x\right)\psi_{s}\left(x\right)$
is the local fermion density operator. A second derivative gives $\partial n/\partial\mu=-\partial^{2}\Omega/\partial\mu^{2}=-\left\langle \left(\int dx\,\hat{n}\left(x\right)\right)^{2}\right\rangle $.
On the other hand, the small momentum limit of the density-density
correlation function is $K^{00}\left(0,\mathbf{q}\rightarrow0\right)=\int dx\int dx^{\prime}K^{00}\left(x,x^{\prime}\right)$,
where $K^{00}\left(x,x^{\prime}\right)=\left\langle \hat{n}\left(x\right)\hat{n}\left(x^{\prime}\right)\right\rangle $
follows from Eq.~\eqref{eq:KAA}. It is straightforward to see this
response function is just $K^{00}\left(0,\mathbf{q}\rightarrow0\right)=-\partial n/\partial\mu$
as defined above. Therefore, the compressibility sum rule is an exact
consequence of a path integral approach \emph{provided no approximations
are made}.

When considering \emph{only} thermodynamics, it is not necessary to
keep track of the vector potential in the self-consistent solution,
and $S_{{\rm eff}}$ can be calculated for $A_{\mu}=0$ and $\bm{\Delta}^{{\rm mf}}\left[0\right]$.
However, when simultaneously considering electrodynamics and thermodynamics
it is important to calculate $S_{{\rm eff}}[A]$ to the same level
of approximation for both quantities. Due to the linear dependence
of both electrodynamic and thermodynamic quantities on the effective
action, any theory studying both quantities, which considers a \emph{consistent}
approximation scheme, will also satisfy the compressibility sum rule.

\paragraph*{Gaussian fluctuations.--}

An exact calculation of $\mathcal{Z}_{{\rm fl}}$ is in general difficult
and is frequently treated at the Gaussian level in the literature.
We similarly consider response at this level: fluctuations $\bm{\eta}$
about the saddle point solution are assumed small and the fluctuation
action is expanded to quadratic order: $S_{\eta}\left[\bm{\Delta}^{{\rm mf}}\left[A\right],A\right]\approx\frac{1}{2}\eta_{a}\widetilde{Q}_{{\rm mf}}^{ab}\eta_{b}$.
The path integral can then be solved exactly; integration of the fluctuation
field $\bm{\eta}$ gives a effective action $S_{{\rm fl}}^{\left(2\right)}=\frac{1}{2}{\rm Tr}\ln\left[\widetilde{Q}_{{\rm mf}}^{ab}\right]$
at the Gaussian level. We emphasize that in the calculation of the
fluctuation response kernel, $K_{{\rm fl}}^{\mu\nu}$, the propagator
$\widetilde{Q}_{{\rm mf}}^{ab}=\widetilde{Q}_{{\rm mf}}^{ab}\left[\bm{\Delta}^{{\rm mf}}\left[A\right],A\right]$
includes dependence on $A_{\mu}$ both explicitly, and through the
mean-field solution. This is in contrast to previous literature which
used the fluctuation propagator $Q_{{\rm mf}}^{ab}$ in Eq.~\eqref{eq:Resp1-1}.

It is clear that setting $A_{\mu}=0$ will reproduce beyond-BCS thermodynamics
found in the literature~\cite{Diener2008,He2015a,Loktev2000,Ohashi2003,Perali2004,Pieri2000,Pieri2005,Taylor2010}.
Similarly, a calculation of $\rho_{s}\sim K^{ii}\left(0,\mathbf{q}\rightarrow0\right)$
will reproduce the bosonic contribution to the superfluid density
found in Refs.~\cite{Taylor2006,Fukushima2007,Taylor2010}. Therefore,
our results reproduce and extend previous explorations of Gaussian
fluctuations, now establishing consistency with the compressibility
sum rule.

\paragraph*{Amplitude and Phase fluctuations.--}

While not explicitly discussed, amplitude fluctuations of the gap
were implicitly included in the compressibility sum rule arguments
presented in this paper. These are often ignored, although they have
been introduced in the literature via an alternative parameterization
of the gap, by writing $\Delta=\rho e^{i2\phi}$, where $\rho=\left|\Delta\right|$
and $2\phi=\arg\Delta$ are respectively the amplitude and phase of
the order parameter. Including amplitude fluctuations by setting $\rho=\rho_{0}+\delta\rho$
and integrating out both $\partial_{\mu}\phi$ and $\delta\rho$ fluctuations
results in a different gauge invariant formulation but one which is
equivalent to the $\bm{\eta}$ fluctuation used above. It should be
noted that while amplitude fluctuations result in a contribution to
electrodynamic (and thermodynamic) response, phase fluctuations alone
are sufficient to restore gauge invariance at both the mean-field
and fluctuation levels. We note, however, that by neglecting amplitude
fluctuations, the compressibility sum rule will be violated and this
violation is apparent even at the mean field level of strict BCS theory.

\paragraph*{Discussion.--}

In this paper we have presented a path integral formulation for superfluids
and superconductors which: (1) allows for a consistent calculation
of (gauge invariant) electrodynamic and thermodynamic response at
any desired level of approximation, and (2) gives the full gauge invariant
response kernel for beyond mean-field physics. The consistency of
our formulation is apparent in the compressibility sum rule which
related electrodynamics and thermodynamics. This sum rule is not satisfied
at the BCS level in the path integral formalism if Gaussian fluctuations
are invoked as in GRGF; instead a consistent treatment involves finding
the saddle point solution in the presence of a vector potential. Our
way of introducing collective mode effects is closer in spirit to
earlier work~\cite{rickayzen} on BCS theory using the Kubo formalism.

We stress an important physical implication of the current scheme.
Within the conventional path integral approach, Gaussian fluctuations
are needed to arrive at gauge invariant electrodynamics. One might
posit that there ought to be fluctuation contributions to thermodynamics.
Specifically, in a neutral superfluid these collective modes would
seem to require power law contributions, say in the specific heat.
We argue here, despite some controversy in the literature~\cite{Yu2009},
including these correction terms in strict BCS theory is unphysical,
as they are inconsistent with the compressibility sum rule.

Within the present formalism, the next level approximation, involving
Gaussian fluctuations then emerges as a true beyond-BCS theory in
which there are inter-related (by the compressibility sum rule) contributions
to both thermodynamics and the electromagnetic response. This beyond-BCS
level of approximation provides a starting point for studying strongly
correlated superfluids. It should be viewed as an alternative to schemes
which build on a correlation self energy and the Ward-Takahashi identity~\cite{ourcompanion}.

This approach provides a promising new route to bench marking beyond-BCS
calculations derived from path integral approaches. There are indications
from the superfluid density at the Gaussian level that possibly unphysical
non-monotonicities appear~\cite{Taylor2010}. These may also be present
when comparing with density correlation functions which are measured
in Bragg scattering experiments. Nevertheless it will be interesting
to look at these higher level (Gaussian) corrections in a variety
of physical contexts, including, for example, their role in topological
\cite{Goryo1998,Goryo1998a,Stone2004,Roy2008,Lutchyn2008,Ojanen2013}
or disordered superfluids~\cite{Kamenev1999}. Quite generally, this
work should be viewed as providing a new paradigm for exploring beyond-BCS
physics using path integral techniques.

\paragraph*{Acknowledgements.--}

We are grateful for illuminating discussions with A. Altland and A.
Kamenev. This work was supported by NSF-DMR-MRSEC 1420709.

\bibliography{GaussFluc}

\clearpage


\section*{Supplementary material}

# Supplemental Material: Correcting inconsistencies in the conventional superfluid path integral scheme

Brandon M. Anderson, Rufus Boyack, Chien-Te Wu, and K. Levin

*James Franck Institute, University of Chicago, Chicago, Illinois 60637, USA*

## PARTITION FUNCTION

We now present a detailed description of the primary result in the main text. We start with the path integral representation of the fermionic partition function in the presence of a non-zero vector potential $A_\mu$:

$$\mathcal{Z}[A] = \int \mathcal{D}[\psi^\dagger, \psi] \, e^{-S_F[\psi^\dagger, \psi, A_\mu]}, \tag{1}$$

where we assume a general fermionic action for a neutral superfluid

$$S_F[\psi^\dagger, \psi, A_\mu] = \int dx \int dy \, \psi_s^\dagger(x) \left(G_0^{-1}[A]\right)_{ss'}(x,y) \psi_{s'}(y) + g \int dx \, \psi_\uparrow^\dagger \psi_\downarrow^\dagger \psi_\uparrow \psi_\downarrow, \tag{2}$$

with an attractive interaction $g > 0$. The fermionic fields $\psi_s^\dagger(x)$ and $\psi_s(x)$ are Grassman numbers to be integrated, whereas $A_\mu$ is an external source that does not fluctuate. The non-interacting inverse Green's function is $G_0^{-1}[A](x,y)$. We assume a neutral $s$-wave superfluid; in a charged superfluid Coulomb interactions can be straightforwardly included at the level of the RPA [1]; the extension to higher order pairings will be left to future work.

We now introduce a Hubbard-Stratonovich (HS) transformation through the identity $1 = \int \mathcal{D}[\boldsymbol{\Delta}] \exp\left[-\int dx \, \frac{|\Delta|^2}{g}\right]$, where the integration measure $\int \mathcal{D}[\boldsymbol{\Delta}]$ is chosen to ensure the integral integrates to unity. The two complex HS fields, represented by $\boldsymbol{\Delta} = (\Delta, \Delta^*)$, will alternatively be denoted through $\boldsymbol{\Delta} = (\Delta_1, \Delta_2)$, where $\Delta_a(x)$, $a = 1,2$ represent the two real fields. These parameterizations are connected by $\Delta_\pm(x) = \Delta_1(x) \pm i\Delta_2(x)$, where the identification $\Delta_- \equiv \Delta$ and $\Delta_+ \equiv \Delta^*$ is consistent with the standard BCS phase convention. We can then write the partition function as

$$\mathcal{Z}[A] = \int \mathcal{D}[\psi^\dagger, \psi, \boldsymbol{\Delta}] \, e^{-S_{F+\text{HS}}[\psi^\dagger, \psi, \boldsymbol{\Delta}, A_\mu]}, \tag{3}$$

where the combined fermionic + Hubbard-Stratonovich action

$$S_{F+\text{HS}}[\psi^\dagger, \psi, \boldsymbol{\Delta}, A_\mu] = \frac{1}{2} \int dx \int dy \, \Psi^\dagger(x) \left(\mathcal{G}_0^{-1}[A](x,y) - \Sigma[\boldsymbol{\Delta}](x,y)\right) \Psi(y) + \int dx \, \frac{|\Delta|^2}{g} \tag{4}$$

is quadratic in the fermionic fields $\psi_s$, $\psi_s^\dagger$, after a shift of integration variables $\Delta \to \Delta - g\psi_\uparrow \psi_\downarrow$ and $\Delta^* \to \Delta^* - g\psi_\uparrow^\dagger \psi_\downarrow^\dagger$. Here $\Psi(x)$ is a conventional Nambu spinor and the single particle Nambu Green's function $\mathcal{G}_0^{-1}[A](x,y)$ takes the standard form [2]. The self-energy is: $\Sigma[\boldsymbol{\Delta}](x,y) = -(\Delta(x)\tau_+ + \Delta^*(x)\tau_-) \delta(x-y)$, with $\tau_+$ ($\tau_-$) a raising (lowering) operator in Nambu space. For notational convenience, the self energy can be expressed using the Nambu Pauli matrices $\tau_{1,2}$ through $\Sigma[\boldsymbol{\Delta}] = -(\Delta_1 \tau_1 + \Delta_2 \tau_2)$.

After applying the HS transformation, the fermionic fields can be integrated out exactly using the standard integration formulas for Grassman numbers. The result is the fermionic partition function expressed exactly as:

$$\mathcal{Z}[A] = \int \mathcal{D}[\boldsymbol{\Delta}] \, e^{-\int dx \, \frac{|\Delta|^2}{g} + \text{Tr} \ln\left[-\mathcal{G}^{-1}[\boldsymbol{\Delta}, A]\right]}. \tag{5}$$

Here $\mathcal{G}^{-1}[\boldsymbol{\Delta}, A](x,y) = \mathcal{G}_0^{-1}[A](x,y) - \Sigma[\boldsymbol{\Delta}](x,y)$ is the inverse Nambu Green's function, which depends on both the HS field $\boldsymbol{\Delta}$, and on the external vector potential $A_\mu$. The $\text{Tr}[\cdot]$ refers to the trace over both Nambu indices, along with a trace (integral) over position indices. This partition function is the starting point in the main text:

$$\mathcal{Z}[A] = \int \mathcal{D}[\boldsymbol{\Delta}] \, e^{-S_{\text{HS}}[\boldsymbol{\Delta}, A]}, \tag{6}$$

where we define the HS action:

$$S_{\text{HS}}[\boldsymbol{\Delta}, A] = \int dx \, \frac{|\Delta|^2}{g} - \text{Tr} \ln\left[-\mathcal{G}^{-1}[\boldsymbol{\Delta}, A]\right]. \tag{7}$$

This is the standard action for a fermionic partition function expressed using the HS transformation.

**Saddle point for $A_\mu \neq 0$**

The arguments in the main text require that the mean-field gap, $\boldsymbol{\Delta}^{\mathrm{mf}}[A]$, be calculated self-consistently in the presence of an arbitrary vector potential $A_\mu \neq 0$. The HS action above can be written as:

$$S_{\mathrm{HS}}[\boldsymbol{\Delta}, A] = \int dx \frac{\Delta_1^2(x) + \Delta_2^2(x)}{g} - \operatorname{Tr}\ln\left[-\mathcal{G}^{-1}[\boldsymbol{\Delta}, A](x, x')\right], \tag{8}$$

where the Nambu Green's function $\mathcal{G}[\boldsymbol{\Delta}, A](x, x')$ is not diagonal in position space. To find the gap equation, one takes the saddle point condition and evaluates at the mean-field solution $\boldsymbol{\Delta}^{\mathrm{mf}}[A](x)$:

$$\left.\frac{\delta S_{\mathrm{HS}}}{\delta \Delta_a(x)}\right|_{\boldsymbol{\Delta}^{\mathrm{mf}}[A]} = 0 = 2\frac{\Delta_a^{\mathrm{mf}}[A](x)}{g} - \operatorname{tr}\left[\mathcal{G}\left[\boldsymbol{\Delta}^{\mathrm{mf}}[A], A\right](x, x)\tau_a\right], \tag{9}$$

and $\operatorname{tr}[\cdot]$ refers to a trace over Nambu indices only. While a solution to this equation is not tractable in general, the non-zero vector potential dependence will be used only to determine how the gap fluctuates with respect to a change in $A_\mu$. At the end of the calculation $A_\mu \to 0$; this allows for the mean-field gap to be calculated for $A_\mu = 0$, and no additional computational complexities arise from this formalism as compared to GRGF.

*Alternative parameterizations*

It is instructive to consider alternative parameterizations of the mean-field degrees of freedom. The above mean-field equations are two simultaneous equations for the gaps $\Delta_1$ and $\Delta_2$. By taking the superposition of these two equations

$$\frac{1}{2}\left(\frac{\delta S_{\mathrm{HS}}}{\delta \Delta_1(x)} \pm i\frac{\delta S_{\mathrm{HS}}}{\delta \Delta_2(x)}\right)_{\boldsymbol{\Delta}^{\mathrm{mf}}[A]} = \frac{\Delta_\pm^{\mathrm{mf}}[A](x)}{g} - \operatorname{tr}\left[\mathcal{G}\left[\boldsymbol{\Delta}^{\mathrm{mf}}[A], A\right](x, x)\tau_\pm\right] \tag{10}$$

we arrive at the well known form of the BCS gap equation for $\Delta^{\mathrm{mf}}$ (and similarly $(\Delta^{\mathrm{mf}})^*$).

Another common parameterization of the mean-field degrees of freedom is $\Delta^{\mathrm{mf}} = \rho e^{i2\phi}$, where $\rho = |\Delta^{\mathrm{mf}}|$, and $2\phi = \arg(\Delta^{\mathrm{mf}})$ are respectively the amplitude and phase of the mean-field gap. (We will suppress the mf superscript on $\rho$ and $\phi$ for notational convenience.) Using this parameterization the Green's function can be expressed as $\mathcal{G}[\boldsymbol{\Delta}^{\mathrm{mf}}[A], A] = \mathcal{G}[\rho[A], \phi[A], A]$. A gauge transformation gives yet another parameterization:

$$\mathcal{G}[\rho[A], \phi[A], A](x, y) \to U(x)\mathcal{G}[\rho[A], 0, A_\mu + \partial_\mu\phi[A]](x, y)U^\dagger(y),$$

where $U(x) = \exp[-i\phi(x)\tau_z]$ is a gauge transformation matrix that "dresses" the vector potential with the mean-field gap. The variational condition can then be expressed as:

$$\frac{\delta S_{\mathrm{mf}}}{\delta \rho(x)} = 0 = 2\frac{\rho(x)}{g} - \operatorname{Tr}[\mathcal{G}[\rho[A], 0, A_\mu + \partial_\mu\phi[A]]\tau_1](x), \tag{11}$$

$$-\partial_\mu \frac{\delta S_{\mathrm{mf}}}{\delta(\partial_\mu\phi(x))} = 0 = \partial_\mu \operatorname{Tr}[\mathcal{G}[\rho[A], 0, A_\mu + \partial_\mu\phi[A]](y, y')\gamma^\mu(y', y, x)], \tag{12}$$

where the current operator, or vertex function, is $\gamma^\mu(y', y, x) = \delta \mathcal{G}_0^{-1}(y', y)/\delta A_\mu(x)$. The first line is the gap equation for the amplitude of the mean-field order parameter. The second equation is the four-divergence of the average current; this gap condition is conceptually understood as the absence of currents in equilibrium. Equations (11) and (12) follow from the functional derivative relations:

$$\frac{\delta S_{\mathrm{mf}}}{\delta A_\mu} = \frac{\delta S_{\mathrm{mf}}}{\delta(\partial_\mu\phi)}, \qquad \partial_\mu \frac{\delta S_{\mathrm{mf}}}{\delta(\partial_\mu\phi)} = -\frac{\delta S_{\mathrm{mf}}}{\delta\phi}.$$

We emphasize these relations are not generic operator identities, but are specific to the form of $S_{\mathrm{mf}}$. However, similar results will hold in related structures below.

Recall that invoking the saddle point condition requires the breaking of a *global* $U(1)$ symmetry (although gauge invariance is of course preserved [3].) If one uses the convention that $\Delta^{\mathrm{mf}}[0]$ is purely real then $\Delta_2^{\mathrm{mf}}[0] = 0$. It follows that the functional derivatives $\delta/\delta\Delta_1^{\mathrm{mf}} = \delta/\delta\rho$ corresponds to amplitude fluctuations, whereas $\delta/\delta\Delta_2^{\mathrm{mf}} = \rho^{-1}\delta/\delta\phi$

corresponds to phase fluctuations. In the limit that $\rho \to 0$, it is known that the amplitude and phase modes decouple. However, for a non-relativistic scalar theory the Bogoliubov and Higgs modes are generically coupled for a finite mean-field amplitude [4, 5]. Therefore, both amplitude and phase fluctuations of the gap are induced by $A_\mu$ (and therefore contribute to response.) However, phase fluctuations are necessary as well as sufficient to ensure gauge invariance of the response kernel.

## Beyond Saddle Point

In order to calculate response beyond saddle point, it is necessary to calculate the partition function at a corresponding level. We generically expand the HS field through

$$\boldsymbol{\Delta} = \boldsymbol{\Delta}^{\mathrm{mf}}[A] + \boldsymbol{\eta}, \tag{13}$$

where $\boldsymbol{\eta} = (\eta_1, \eta_2)$ is a fluctuation around the mean-field solution. (Similar to $\boldsymbol{\Delta}$, we can also use the $\eta_\pm = \eta_1 \pm i\eta_2$ parameterization.) We do not yet make any assumptions about the size of $\boldsymbol{\eta}$, but we note it is a dynamic variable so it does not contain dependence on either the vector potential $A_\mu$, or the mean-field solution $\boldsymbol{\Delta}^{\mathrm{mf}}[A]$. The partition function can be exactly factorized as $\mathcal{Z}[A] = \mathcal{Z}_{\mathrm{mf}}\left[\boldsymbol{\Delta}^{\mathrm{mf}}[A], A\right] \mathcal{Z}_{\mathrm{fl}}\left[\boldsymbol{\Delta}^{\mathrm{mf}}[A], A\right]$, where

$$\mathcal{Z}_{\mathrm{fl}}\left[\boldsymbol{\Delta}^{\mathrm{mf}}[A], A\right] = \int \mathcal{D}[\boldsymbol{\eta}] \, e^{-S_\eta\left[\boldsymbol{\Delta}^{\mathrm{mf}}[A], A, \boldsymbol{\eta}\right]} \tag{14}$$

is the partition function describing beyond-saddle-point fluctuations, and

$$S_\eta\left[\boldsymbol{\Delta}^{\mathrm{mf}}[A], A, \boldsymbol{\eta}\right] = S_{\mathrm{HS}}\left[\boldsymbol{\Delta}^{\mathrm{mf}}[A] + \boldsymbol{\eta}, A\right] - S_{\mathrm{HS}}\left[\boldsymbol{\Delta}^{\mathrm{mf}}[A], A\right] \tag{15}$$

is the action of a fluctuation $\boldsymbol{\eta}$ around the mean-field solution. While this action is not assumed to be small, the lowest order correction is $\mathcal{O}\left(\boldsymbol{\eta}^2\right)$, since the term linear in $\boldsymbol{\eta}$ exactly vanishes by the saddle point condition. Note also that under a gauge transformation $A_\mu \to A_\mu + \partial_\mu \alpha$, the mean-field gap transforms as $\boldsymbol{\Delta}^{\mathrm{mf}}[A] \to e^{2i\alpha} \boldsymbol{\Delta}^{\mathrm{mf}}[A]$. Therefore, after a change of variables in $\boldsymbol{\eta}$, both $\mathcal{Z}_{\mathrm{mf}}$ and $\mathcal{Z}_{\mathrm{fl}}$ are gauge invariant, provided the mean-field solution is properly gauge transformed.

If the partition function is desired at the Gaussian fluctuation level, it is straightforward to see that the action $S_\eta$ can be expanded to second order to give

$$S_\eta\left[\boldsymbol{\Delta}^{\mathrm{mf}}[A], A, \boldsymbol{\eta}\right] \approx \frac{1}{2} \int dx \int dy \sum_{ab} \eta_a(x) \widetilde{Q}_{\mathrm{mf}}^{ab}(x, y) \eta_b(y), \tag{16}$$

where $\widetilde{Q}_{\mathrm{mf}}^{ab} = \widetilde{Q}_{\mathrm{mf}}^{ab}\left[\boldsymbol{\Delta}^{\mathrm{mf}}[A], A\right]$ is the conventional mean-field propagator, except calculated at a gap $\boldsymbol{\Delta}^{\mathrm{mf}}[A]$, and nonzero $A_\mu$. We can then integrate out the fluctuation field $\boldsymbol{\eta}$ explicitly giving

$$\mathcal{Z}_{\mathrm{fl}}\left[\boldsymbol{\Delta}^{\mathrm{mf}}[A], A\right] \approx e^{-S_{\mathrm{fl}}^{(2)}\left[\boldsymbol{\Delta}^{\mathrm{mf}}[A], A\right]}, \tag{17}$$

where the well-known Gaussian fluctuation action

$$S_{\mathrm{fl}}^{(2)}\left[\boldsymbol{\Delta}^{\mathrm{mf}}[A], A\right] = \frac{1}{2} \mathrm{Tr} \ln \left[\widetilde{Q}_{\mathrm{mf}}^{ab}\right] \tag{18}$$

depends on the mean-field solution $\boldsymbol{\Delta}^{\mathrm{mf}}[A]$ and the vector potential $A_\mu$.

## GENERAL RESPONSE KERNEL

In the main text, we are interested in the gauge invariant response kernels for both the mean-field and beyond mean-field response. We note that a general response function is found from expanding the effective action $S_{\mathrm{eff}}[A] = -\ln \mathcal{Z}[A]$ to second order in the vector potential $A_\mu$ [6]. From the above parameterization of the mean-field/fluctuation partition function, we see that the total effective action is then linear in the saddle-point and fluctuation actions: $S_{\mathrm{eff}}[A] = S_{\mathrm{mf}}[A] + S_{\mathrm{fl}}[A]$. As a result, the response kernel will also be linear in the successive approximations for $\mathcal{Z}[A]$. We write out a generic effective action as $S_i[A] = -\ln \mathcal{Z}_i[A]$, where $i$ can be eff, mf, or fl above, and expand to second order in $A_\mu$ to obtain the general response kernel, defined through:

$$K_i^{\mu\nu}(x, x') = \left.\frac{\delta^2 S_i[A]}{\delta A_\mu(x) \delta A_\nu(x')}\right|_{A \to 0}. \tag{19}$$

In what follows we will show this has the form presented in Eq. (4) of the main text.

### Functional chain rule

By invoking the saddle point condition, the partition function depends on the vector potential through the mean-field solution, so $\mathcal{Z}_i[A] = \mathcal{Z}_i[\boldsymbol{\Delta}^{\mathrm{mf}}[A], A]$, where $\boldsymbol{\Delta}^{\mathrm{mf}}[A]$ is a self-consistent mean-field gap. When taking the second order functional derivatives of $S_i$, an additional contribution appears due to a "functional chain rule" [2] arising from the dependence of the mean-field gap on the vector potential. These contributions will emerge as follows:

$$\frac{\delta S_i[\boldsymbol{\Delta}^{\mathrm{mf}}[A], A]}{\delta A_\mu(x)} = \left(\frac{\delta S_i[\boldsymbol{\Delta}, A]}{\delta A_\mu(x)}\right)_{\boldsymbol{\Delta}^{\mathrm{mf}}[A]} + \int dy \frac{\delta \Delta_a^{\mathrm{mf}}[A](y)}{\delta A_\mu(x)} \left(\frac{\delta S_i[\boldsymbol{\Delta}, A]}{\delta \Delta_a(y)}\right)_{\boldsymbol{\Delta}^{\mathrm{mf}}[A]}. \tag{20}$$

In this expression and below, terms such as $(\delta S_i[\boldsymbol{\Delta}, A]/\delta A_\mu)_{\boldsymbol{\Delta}^{\mathrm{mf}}[A]}$ or $(\delta S_i[\boldsymbol{\Delta}, A]/\delta \Delta_a)_{\boldsymbol{\Delta}^{\mathrm{mf}}[A]}$ are evaluated by differentiating only the explicit $A_\mu$ or $\Delta_a$ dependence, and then evaluating the result at $\boldsymbol{\Delta} = \boldsymbol{\Delta}^{\mathrm{mf}}[A]$. In this way, these can be interpreted as "partial functional derivatives." We have not yet taken the $A_\mu \to 0$ limit; taking the second functional derivative, a similar procedure results in:

$$\frac{\delta^2 S_i[\boldsymbol{\Delta}^{\mathrm{mf}}[A], A]}{\delta A_\mu(x) \delta A_\nu(x')} = \left(\frac{\delta^2 S_i[\boldsymbol{\Delta}, A]}{\delta A_\mu(x) \delta A_\nu(x')}\right)_{\boldsymbol{\Delta}^{\mathrm{mf}}[A]} + \int dy \int dy' \frac{\delta \Delta_a^{\mathrm{mf}}[A](y)}{\delta A_\mu(x)} \left(\frac{\delta^2 S_i[\boldsymbol{\Delta}, A]}{\delta \Delta_a(y) \delta \Delta_b(y')}\right)_{\boldsymbol{\Delta}^{\mathrm{mf}}[A]} \frac{\delta \Delta_b^{\mathrm{mf}}[A](y')}{\delta A_\nu(x')}$$
$$+ \int dy \frac{\delta \Delta_a^{\mathrm{mf}}[A](y)}{\delta A_\mu(x)} \left(\frac{\delta^2 S_i[\boldsymbol{\Delta}, A]}{\delta \Delta_a(y) \delta A_\nu(x')}\right)_{\boldsymbol{\Delta}^{\mathrm{mf}}[A]} + \int dy \left(\frac{\delta^2 S_i[\boldsymbol{\Delta}, A]}{\delta A_\mu(x) \delta \Delta_a(y)}\right)_{\boldsymbol{\Delta}^{\mathrm{mf}}[A]} \frac{\delta \Delta_a^{\mathrm{mf}}[A](y)}{\delta A_\nu(x')}$$
$$+ \int dy \left(\frac{\delta S_i[\boldsymbol{\Delta}, A]}{\delta \Delta_a(y)}\right)_{\boldsymbol{\Delta}^{\mathrm{mf}}[A]} \frac{\delta^2 \Delta_a^{\mathrm{mf}}[A](y)}{\delta A_\mu(x) \delta A_\nu(x')}. \tag{21}$$

The response is then found by applying the $A_\mu \to 0$ limit to the above equation. The result is Eq. (4) in the main text for $i = \mathrm{mf}$, along with the functional equivalent for $i = \mathrm{fl}$.

To reproduce Eq. (6) of the main text, it is convenient to define a set of two-point correlation functions

$$\mathcal{Q}_i^{\alpha\beta}(x, x') = \left.\frac{\delta^2 S_i[\boldsymbol{\Delta}^{\mathrm{mf}}, A]}{\delta \mathcal{A}_\alpha(x) \delta \mathcal{A}_\beta(x')}\right|_{A \to 0}, \tag{22}$$

with a generalized response vector $\mathcal{A}_\alpha = (\Delta_1^{\mathrm{mf}}, \Delta_2^{\mathrm{mf}}, A_\mu)$ that combines simultaneous gap and vector potential fluctuations. Due to the order of limits, we take $\boldsymbol{\Delta}^{\mathrm{mf}} \equiv \boldsymbol{\Delta}^{\mathrm{mf}}[0]$, and do not include any chain rule terms arising from $\boldsymbol{\Delta}^{\mathrm{mf}}[A]$. From here, the two-point functions in the main text are $K_{0,i}^{\mu\nu} = \mathcal{Q}_i^{\mu\nu}$, $Q_i^{\mu a} = \mathcal{Q}_i^{\mu a}$, and $Q_i^{ab} = \mathcal{Q}_i^{ab}$. It is also helpful to define the first and second order gap fluctuations

$$\Pi_a^\mu(x, x') = \left.\frac{\delta \Delta_a^{\mathrm{mf}}[A](x')}{\delta A_\mu(x)}\right|_{A \to 0},$$
$$\Xi_a^{\mu\nu}(x, x', x'') = \left.\frac{\delta^2 \Delta_a^{\mathrm{mf}}[A](x')}{\delta A_\mu(x) \delta A_\nu(x'')}\right|_{A \to 0}, \tag{23}$$

whose form will be calculated explicitly below. Using these expressions, the full gauge-invariant response kernel for an action $S_i[\boldsymbol{\Delta}^{\mathrm{mf}}[A], A]$ is

$$K_i^{\mu\nu}(x, x') = \left.\frac{\delta^2 S_i[\boldsymbol{\Delta}^{\mathrm{mf}}[A], A]}{\delta A_\mu(x) \delta A_\nu(x')}\right|_{A \to 0},$$
$$= K_{0,i}^{\mu\nu}(x, x') + \int dy \int dy' \Pi_a^\mu(x, y) Q_i^{ab}(y, y') \Pi_b^\nu(x', y')$$
$$+ \int dy \left(\Pi_a^\mu(x, y) Q_i^{a\nu}(y, x') + Q_i^{\mu a}(x, y) \Pi_a^\nu(x', y)\right)$$
$$+ \int dy \frac{\delta S_i[\boldsymbol{\Delta}^{\mathrm{mf}}, 0]}{\delta \Delta_a^{\mathrm{mf}}(y)} \Xi_a^{\mu\nu}(x, y, x'). \tag{24}$$

Since the $i$-index is general, this equation produces the mean-field response as derived in the main text for $i = \mathrm{mf}$, when the stationary condition causes the last line to vanish in Eq. 24. For a generic fluctuation action, the stationary condition may not be satisfied, and the third line may be non-zero. In this case, the second order gap fluctuation will contribute and this term is key to maintaining gauge invariance for non-stationary actions. This will be explicitly shown below.

### Gap equation and collective mode propagators

We now use the gap equation, or saddle point condition, to calculate the collective mode terms $\Pi_a^\mu(x,x') = \delta\Delta_a^{\mathrm{mf}}[A](x')/\delta A_\mu(x)|_{A\to 0}$. We start by using the saddle-point condition for the gap equation in the presence of $A_\mu \neq 0$:

$$0 = \frac{\delta S_{\mathrm{mf}}\left[\boldsymbol{\Delta}^{\mathrm{mf}}[A], A\right]}{\delta \Delta_a^{\mathrm{mf}}[A](x')}. \tag{25}$$

We then follow the procedure above and take the functional derivative with respect to $\delta/\delta A_\mu(x)$, invoking the functional chain rule:

$$0 = \frac{\delta}{\delta A_\mu(x)}\left(\frac{\delta S_{\mathrm{mf}}[\boldsymbol{\Delta},A]}{\delta \Delta_a(x')}\right)_{\boldsymbol{\Delta}^{\mathrm{mf}}[A]},$$
$$= \left(\frac{\delta^2 S_{\mathrm{mf}}[\boldsymbol{\Delta},A]}{\delta A_\mu(x)\,\delta \Delta_a(x')}\right)_{\boldsymbol{\Delta}^{\mathrm{mf}}[A]} + \int dy\, \frac{\delta \Delta_b^{\mathrm{mf}}[A](y)}{\delta A_\mu(x)}\left(\frac{\delta^2 S_{\mathrm{mf}}[\boldsymbol{\Delta},A]}{\delta \Delta_b(y)\,\delta \Delta_a(x')}\right)_{\boldsymbol{\Delta}^{\mathrm{mf}}[A]}. \tag{26}$$

After taking $A_\mu \to 0$ limit, this is expressed using the notation introduced above as:

$$\int dy\, \Pi_b^\mu(x,y)\, Q_{\mathrm{mf}}^{ba}(y,x') = -Q_{\mathrm{mf}}^{\mu a}(x,x'). \tag{27}$$

This is a linear equation of the form $M\mathbf{x} = \mathbf{b}$, where $Q_{\mathrm{mf}}^{ab}(x,y)$, $\Pi_b^\mu(x,y)$, and $Q_i^{\mu a}(x,x')$ respectively take the role of the matrix $M$, $\mathbf{x}$, and $\mathbf{b}$. Therefore, provided $Q_{\mathrm{mf}}^{ab}(x,y)$ is invertible, we can write:

$$\Pi_a^\mu(x,x') = -\int dy\, [Q_{\mathrm{mf}}]_{ab}^{-1}(x',y)\, Q_{\mathrm{mf}}^{b\mu}(y,x), \tag{28}$$

where $[Q_{\mathrm{mf}}]_{ab}^{-1}(x,y)$ is the inverse of the bosonic fluctuation propagator with respect to both Nambu matrix elements $a,b$ and position matrix elements $x,y$, and for notational clarity we have used the relation $Q_{\mathrm{mf}}^{b\mu}(y,x) = Q_{\mathrm{mf}}^{\mu b}(x,y)$.

We note that the inverse of the two-point function $Q_{\mathrm{mf}}^{ab}(x,y)$, which can be interpreted as a bosonic propagator [7, 8], will have a pole corresponding to the gap equation [9–11]. The collective mode terms can therefore be calculated using the textbook resolvent method [2] for finding Green's functions. For a translationally invariant system, we convert to momentum space, and the inverse propagator then becomes $\left[Q_{\mathrm{mf}}^{ab}(q)\right]^{-1}\delta_{-qq'}$. The inverse of the propagator is just the inverse of the $2\times 2$ matrix $Q_{\mathrm{mf}}^{ab}(q)$ for each $q_\mu \neq 0$, whereas $q_\mu \to 0$ gives the BCS gap equation. Converting $\left[Q_{\mathrm{mf}}^{ab}(q)\right]^{-1}$ to position space gives the inverse propagator $[Q_{\mathrm{mf}}]_{ab}^{-1}(x,y)$.

To understand the $q_\mu \to 0$ pole of the bosonic propagator, note that the collective modes must be singular in that limit. This should not be surprising since it is well known [9–11] that the collective modes satisfy $q_\mu \Pi_a^\mu(q) = 2i\epsilon_{ab}\Delta_b$ for all $q_\mu$. Therefore, as $q_\mu \to 0$, $\Pi_a^\mu \sim q^\mu/q^2$ must have a simple pole to maintain a finite contraction at $q_\mu = 0$. While the collective modes for finite $A_\mu$ are in general complicated, taking $A_\mu \to 0$ in a translationally invariant system produces the well known result:

$$\Pi_a^\mu(q) = -[Q_{\mathrm{mf}}]_{ab}^{-1}(q)\, Q_{\mathrm{mf}}^{b\mu}(q). \tag{29}$$

In this case, the simple pole in $\Pi_a^\mu(q)$ is directly inherited from the bosonic propagator $[Q_{\mathrm{mf}}]_{ab}^{-1}(q)$.

### Second order fluctuation of $\Delta^{\mathrm{mf}}[A]$

We can calculate the second order fluctuation of the gap, denoted $\Xi_a^{\mu\nu}(x,x',x'')$, in a manner similar to above. The general expression involves three derivatives of the saddle point action with respect to $A_\mu$ or $\Delta_a^{\mathrm{mf}}$. The calculation is a straightforward extension of the one in the previous section. The result is given by:

$$\Xi_a^{\mu\nu}(x,x',x'') \equiv \left.\frac{\delta^2 \Delta_a^{\mathrm{mf}}[A](x')}{\delta A_\mu(x)\,\delta A_\nu(x'')}\right|_{A\to 0}, \tag{30}$$
$$= -\int d\xi\, [Q_{\mathrm{mf}}]_{aa'}^{-1}(x',\xi)\left(Q_{\mathrm{mf}}^{\mu a'\nu}(x,\xi,x'') + \int dy\, \Pi_b^\mu(x,y)\, Q_{\mathrm{mf}}^{ba'\nu}(y,\xi,x'')\right)$$
$$- \int d\xi\, [Q_{\mathrm{mf}}]_{aa'}^{-1}(x',\xi)\int dy'\left(Q_{\mathrm{mf}}^{\mu a'b}(x,\xi,y') + \int dy\, \Pi_c^\mu(x,y)\, Q_{\mathrm{mf}}^{ca'b}(y,\xi,y')\right)\Pi_b^\nu(x'',y'),$$

where we now define the three-point functions

$$Q_{\mathrm{mf}}^{\alpha\beta\gamma}\left(x,x',x''\right) = \left.\frac{\delta^3 S_{\mathrm{mf}}\left[\boldsymbol{\Delta}^{\mathrm{mf}},A\right]}{\delta\mathcal{A}_\alpha\left(x\right)\delta\mathcal{A}_\beta\left(x'\right)\delta\mathcal{A}_\gamma\left(x''\right)}\right|_{A\to 0}, \tag{31}$$

similarly to the two-point correlation functions in Eq (22).

## CHECKING GAUGE INVARIANCE

We now confirm gauge invariance for the response kernel in Eq. (24) for an arbitrary action $S_i\left[\boldsymbol{\Delta}^{\mathrm{mf}}\left[A\right],A\right]$. Before we do this, we note that by construction, the effective action is gauge invariant. Therefore, a proper $A_\mu \to 0$ expansion will automatically maintain gauge invariance. Any violation of gauge invariance must be a result of an improperly calculated functional expansion. In linear response theory, the condition

$$\partial_\mu^x K_i^{\mu\nu}\left(x,x'\right) = 0$$

is necessary and sufficient for gauge invariance of a given response kernel [6].

To check the calculated response kernel in Eq. (24) satisfies the gauge invariant condition above, it is helpful to write:

$$K_i^{\mu\nu}\left(x,x'\right) = K_{0,i}^{\mu\nu}\left(x,x'\right) + \int dy\,\Pi_a^\mu\left(x,y\right)Q_i^{a\nu}\left(y,x'\right)$$
$$+ \int dy'\,\left(Q_i^{\mu b}\left(x,y\right) + \int dy\,\Pi_a^\mu\left(x,y\right)Q_i^{ab}\left(y,y'\right)\right)\Pi_b^\nu\left(x',y'\right)$$
$$+ \int dy\,\frac{\delta S_i}{\delta\Delta_a^{\mathrm{mf}}\left(y\right)}\Xi_a^{\mu\nu}\left(x,y,x'\right). \tag{32}$$

For the $i = \mathrm{mf}$ action only, the second and third lines vanish explicitly without any contraction. If the saddle point action is not considered, calculations of the contractions is technically involved. We leave the details to a later section. We now present the line-by-line results of the contractions of the full response above:

$$\partial_\mu^x\left(K_{0,i}^{\mu\nu}\left(x,x'\right) + \int dy\,\Pi_a^\mu\left(x,y\right)Q_i^{a\nu}\left(y,x'\right)\right) = 0, \tag{33}$$

$$\partial_\mu^x\left(Q_i^{\mu b}\left(x,y'\right) + \int dy\,\Pi_a^\mu\left(x,y\right)Q_i^{ab}\left(y,y'\right)\right) = -2\frac{\delta S_i}{\delta\Delta_a^{\mathrm{mf}}\left(y'\right)}\epsilon_{ab}\delta\left(x-y'\right), \tag{34}$$

$$\partial_\mu^x\Xi_a^{\mu\nu}\left(x,y,x'\right) = 2\epsilon_{ab}\delta\left(x-y\right)\Pi_b^\nu\left(y,x'\right). \tag{35}$$

Substituting these relations into Eq. (32) gives the contraction of the response kernel:

$$\partial_\mu K_i^{\mu\nu}\left(x,x'\right) = \int dy'\,\left(-2\frac{\delta S_i}{\delta\Delta_a^{\mathrm{mf}}\left(y'\right)}\epsilon_{ab}\delta\left(x-y'\right)\right)\Pi_b^\nu\left(y',x'\right)$$
$$+ \int dy\,\frac{\delta S_i}{\delta\Delta_a^{\mathrm{mf}}\left(y\right)}\left(2\epsilon_{ab}\delta\left(x-y\right)\Pi_b^\nu\left(y,x'\right)\right),$$
$$= 0.$$

Therefore, the contraction of the response kernel vanishes, and we conclude the general response kernel Eq. (24) is gauge invariant, provided the relations in Eqs. (33)-(35) hold.

### Useful contraction formulas

We will now derive these contraction relations systematically. In order to derive Eqs. (33)-(35), we will first consider a set of formulas for the contraction of a general $n$-point correlation function, as calculated from functional derivatives of the gauge invariant action with respect to either $A_\mu$, or $\Delta_a^{\mathrm{mf}}\left[0\right]$. The desired contraction relations will immediately follow from limiting cases.

The relevant correlation functions, such as those in Eq. (33), are all partial derivatives calculated with $\boldsymbol{\Delta} = \boldsymbol{\Delta}^{\mathrm{mf}}\left[0\right]$. Since the gap in these functions is independent of $A_\mu$, the functional chain rule should not invoked.

In order to simplify notation, *in this section only,* we will drop the mf superscript on the mean-field gaps, so that $\boldsymbol{\Delta} = \boldsymbol{\Delta}^{\mathrm{mf}}\left[0\right]$.

### Generic contracted n-point function

All correlation functions of interest will be of the form

$$\mathcal{Q}_i^{\alpha_1\ldots\alpha_n}(x_1\ldots x_n) \equiv \left.\frac{\delta^n S_i[\mathbf{\Delta}, A]}{\delta\mathcal{A}_1\ldots\delta\mathcal{A}_n}\right|_{A\to 0}, \tag{36}$$

where $\mathcal{A}_m = \mathcal{A}_{\alpha_m}(x_m)$ is a generalized response vector that contains both the gap and vector potential, as defined below Eq. (22). Since the gauge invariance condition requires the contraction of an index, it will be helpful to calculate a series of contraction formulas:

$$\partial_\mu^x \frac{\delta}{\delta A_\mu(x)} \mathcal{Q}_i^{\{\alpha_n\}}(\{x_n\}), \tag{37}$$

where the shorthand $\{x_n\} = x_1\ldots x_n$ and $\{\alpha_n\} = \alpha_1\ldots\alpha_n$. In order to calculate these contraction formulas, first note that all functional derivatives with respect to $\Delta_{a_n}$ and $A_{\mu_n}$ commute. We also assume $x_m \neq x$ for any $m$, so the partial derivative $\partial_x$ can be commuted through the functional derivatives. The contracted correlation function is equivalently expressed as:

$$\partial_\mu^x \frac{\delta}{\delta A_\mu(x)} \mathcal{Q}_i^{\{\alpha_n\}}(\{x_n\}) = \frac{\delta^n}{\delta\mathcal{A}_1\ldots\delta\mathcal{A}_n}\left(\partial_\mu^x \frac{\delta S_i[\mathbf{\Delta}, A]}{\delta A_\mu(x)}\right). \tag{38}$$

In the path integral, the actions $S_i[\mathbf{\Delta}, A] = S_i[\rho, \phi, A]$, are manifestly gauge invariant. Therefore, we can apply a gauge transformation $\Delta = \rho e^{2i\phi} \to \rho$ and $A_\mu \to A_\mu + \partial_\mu\phi$ to move all phase dependence of the order parameter into a "dressed" vector potential $\tilde{A}_\mu = A_\mu + \partial_\mu\phi$ [1, 12]. Under this transformation, the gauge invariant action must not change, and the three parameterizations of the action

$$S_i[\mathbf{\Delta}, A] = S_i[\rho, \phi, A] = S_i[\rho, 0, A + \partial\phi] \tag{39}$$

are equivalent for both a saddle point and a generic fluctuation action. Returning to the contraction formulas, we see the above gauge invariant reparameterizations of the action $S_i$ imply

$$\partial_\mu^x \frac{\delta S_i[\mathbf{\Delta}, A]}{\delta A_\mu(x)} = \partial_\mu^x \frac{\delta S_i[\rho, 0, A + \partial\phi]}{\delta(\partial_\mu^x \phi(x))} = -\frac{\delta S_i[\rho, 0, A + \partial\phi]}{\delta\phi(x)} = -\frac{\delta S_i[\rho, \phi, A]}{\delta\phi(x)}. \tag{40}$$

The first equality follows from gauge invariance and the functional chain rule. The second equality is a result of the functional derivative identity $\frac{\delta F[\partial\phi]}{\delta\phi} = -\partial\frac{\delta F[\partial\phi]}{\delta(\partial\phi)}$, which is true if $F[\partial\phi]$ is a functional of *only* $\partial_\mu\phi$ (and not $\phi$ and $\partial_\mu\phi$ simultaneously.) This condition is satisfied when the action is expressed as $S_i[\rho, 0, A + \partial\phi]$ so the vector potential is dressed with the mean-field phase. The last equality used the property of gauge invariance in the choice of the parameterization of the action in Eq. (39).

The contraction formula can then be expressed as:

$$\partial_\mu^x \frac{\delta}{\delta A_\mu(x)} \mathcal{Q}_i^{\alpha_1\ldots\alpha_n}(x_1\ldots x_n) = -\frac{\delta^n}{\delta\mathcal{A}_1\ldots\delta\mathcal{A}_n}\frac{\delta S_i}{\delta\phi} = -\frac{\delta^n}{\delta\mathcal{A}_1\ldots\delta\mathcal{A}_n}\frac{\delta\Delta_a}{\delta\phi}\frac{\delta S_i}{\delta\Delta_a}, \tag{41}$$

where we henceforth suppress the functional argument on the action $S_i$. While the functional derivatives $\delta/\delta\mathcal{A}_\alpha$ commute between themselves, note that $\delta/\delta\Delta_a^{\mathrm{mf}}$ and $\delta/\delta\phi$ do not commute, nor do they commute with $\delta\Delta_a/\delta\phi$. This will lead to a product rule with respect to the $n$ functional derivatives of $\mathcal{A}_\alpha$. Since $\delta\Delta_a/\delta\phi = 2\epsilon_{ab}\Delta_b$ is linear in the gap $\Delta_b$, only terms with a single functional derivative applied to $\delta\Delta_a/\delta\phi$ will contribute:

$$\frac{\delta^n}{\delta\mathcal{A}_1\ldots\delta\mathcal{A}_n}\frac{\delta\Delta_a}{\delta\phi}\frac{\delta S_i}{\delta\Delta_a} = \frac{\delta\Delta_a}{\delta\phi}\frac{\delta^{n+1} S_i}{\delta\mathcal{A}_1\ldots\delta\mathcal{A}_n\delta\Delta_a} + \sum_{m=1}^n\left(\frac{\delta}{\delta\mathcal{A}_{\alpha_m}}\frac{\delta\Delta_a}{\delta\phi}\right)\frac{\delta^n S_i}{\prod_{k\neq m}\delta\mathcal{A}_k\delta\Delta_a},$$

$$= \frac{\delta\mathcal{Q}_i^{\{\alpha_n\}}(\{x_n\})}{\delta\phi(x)} + 2\sum_{m=1}^n \epsilon_{ab}\delta_{b\alpha_m}\delta(x - x_m)\mathcal{Q}_i^{\{\alpha_{n\neq m}, a\}}(\{x_{n\neq m}, x\}). \tag{42}$$

The contraction of an arbitrary correlation function is therefore given by

$$\partial_\mu^x \frac{\delta}{\delta A_\mu(x)} \mathcal{Q}_i^{\{\alpha_n\}}(\{x_n\}) = -\frac{\delta\mathcal{Q}_i^{\{\alpha_n\}}(\{x_n\})}{\delta\phi} - 2\sum_{m=1}^n \epsilon_{ab}\delta_{b\alpha_m}\delta(x - x_m)\mathcal{Q}_i^{\{\alpha_{n\neq m}, a\}}(\{x_{n\neq m}, x\}), \tag{43}$$

for any set of arbitrary $n$-point correlation functions calculated from an effective action $S_i[\mathbf{\Delta}, A]$.

*Contractions in two-point functions*

This contraction formula immediately produces two useful contractions: $\partial_\mu K_{0,i}^{\mu\nu}$ and $\partial_\mu Q_i^{\mu a}$; both contractions correspond to the operator $\partial_\mu \frac{\delta}{\delta A_\mu}$ applied to the one-point correlation function $\mathcal{Q}_i^\alpha(y)$. Directly applying the above formula gives:

$$\partial_\mu^x \frac{\delta}{\delta A_\mu(x)} \mathcal{Q}_i^\alpha(y) = -\frac{\delta \mathcal{Q}_i^\alpha(y)}{\delta \phi} - 2\epsilon_{bc}\delta_{c\alpha} \mathcal{Q}_i^b(x) \delta(x-y). \tag{44}$$

The bubble response kernel is $\alpha = \nu$, and the $\delta_{c\nu} = 0$ delta function causes the term to the right to vanish. For the $Q_i^{\mu a}$ term, the delta function does not vanish. The two contractions are generically expressed as:

$$\partial_\mu^x K_{0,i}^{\mu\nu}(x,y) = -\frac{\delta \mathcal{Q}_i^\nu(y)}{\delta \phi(x)}, \tag{45}$$

$$\partial_\mu^x Q_i^{\mu a}(x,y) = -\frac{\delta \mathcal{Q}_i^a(y)}{\delta \phi(x)} - 2\epsilon_{ba} \mathcal{Q}_i^b(x) \delta(x-y). \tag{46}$$

If $i = \text{mf}$, the $\mathcal{Q}_i^b(x)$ term on the second line corresponds to the saddle point condition, and therefore vanishes. Otherwise, this term needs to be kept, and it will be shown that this is an important contribution in maintaining gauge invariance of $K^{\mu\nu}$ for non-stationary actions $\delta S_i/\delta \Delta_a \neq 0$.

Of special interest is the contraction $\partial_\mu^x Q_{\text{mf}}^{\mu a}$, which evaluates to:

$$\partial_\mu^x Q_{\text{mf}}^{\mu a}(x,x') = -2\epsilon_{bc} \Delta_b(x) Q_{\text{mf}}^{ca}(x,x') \tag{47}$$

after using the functional identity

$$\frac{\delta}{\delta \phi} = \frac{\delta \Delta_a}{\delta \phi} \frac{\delta}{\delta \Delta_a} = 2\epsilon_{ab}\Delta_a \frac{\delta}{\delta \Delta_b}. \tag{48}$$

Using Eq. (47) and Eq. (26), the contraction $\partial_\mu^x \Pi_a^\mu(x,x')$ over the collective mode vertices follows straightforwardly:

$$\partial_\mu^x \Pi_a^\mu(x,x') = -\int dy\, [Q_{\text{mf}}]_{ab}^{-1}(x',y) \left(\partial_\mu^x Q_{\text{mf}}^{b\mu}(y,x)\right),$$
$$= 2\epsilon_{a'a}\Delta_{a'}(x)\delta(x-x'). \tag{49}$$

A related set of contraction formulas will also be helpful

$$\partial_\mu \int dy\, \Pi_a^\mu(x,y) \mathcal{Q}_i^{\{a\alpha_n\}}(y,\{x_n\}) = \int dy\, (2\epsilon_{ab}\Delta_b(x)\delta(x-y)) \mathcal{Q}_i^{\{a\alpha_n\}}(y,\{x_n\}),$$
$$= 2\Delta_a(x) \epsilon_{ab} \frac{\delta}{\delta \Delta_b(x)} \mathcal{Q}_i^{\{\alpha_n\}}(\{x_n\}),$$
$$= \frac{\delta}{\delta \phi(x)} \mathcal{Q}_i^{\{\alpha_n\}}(\{x_n\}). \tag{50}$$

This shows that the contraction of a collective mode propagator acts as a $\phi$-derivative, but with the opposite sign as a generic correlation function $\mathcal{Q}^{\{\alpha_n\}}(\{x_n\})$. On a physical level, this can be interpreted as an equivalency between "pure gauge" fluctuations and phase fluctuations of the order parameter. From a similar standpoint, we see that amplitude collective modes are not responsible for the restoration of gauge invariance, even though they contribute to response. Combined with Eq. (43) above, we find the functional identity:

$$\partial_\mu^x \left( \mathcal{Q}_i^{\{\mu\alpha_n\}}(x,\{x_n\}) + \int dy\, \Pi_a^\mu(x,y) \mathcal{Q}_i^{\{a\alpha_n\}}(y,\{x_n\}) \right) = -2 \sum_{m=1}^n \epsilon_{ab}\delta_{b\alpha_m}\delta(x-x_m) \mathcal{Q}_i^{\{\alpha_{n\neq m},a\}}(\{x_{n\neq m},x\}). \tag{51}$$

*Specific contractions in the gauge-invariant response kernel*

These results are now sufficient to show the full response is gauge invariant, and $\partial_\mu K_i^{\mu\nu} = 0$. We examine the three lines (Eq. (33)-(35) above) in the contracted response. The contraction of the first line

$$\partial_\mu^x \left( K_{0,i}^{\mu\nu}(x,x') + \int dy\, \Pi_a^\mu(x,y) Q_i^{a\nu}(y,x') \right) = -\frac{\delta}{\delta \phi(x)} \mathcal{Q}_i^\nu(x') + \frac{\delta}{\delta \phi(x)} \mathcal{Q}_i^\nu(x') = 0, \tag{52}$$

vanishes explicitly. When contracting the second line, the term in the parentheses does not vanish due to the non-commutativity of $\delta/\delta\phi$ and $\delta/\delta\Delta_a$:

$$\partial_\mu^x \left( Q_i^{\mu a}(x,y') + \int dy\, \Pi_a^\mu(x,y) Q_i^{ab}(y,y') \right) = -\left( \frac{\delta}{\delta\phi(x)} \mathcal{Q}_i^a(y') + 2\sum_{m=1}^n \epsilon_{ba} \mathcal{Q}_i^b(y') \delta(x-y') \right) + \frac{\delta}{\delta\phi(x)} \mathcal{Q}_i^a(y'),$$
$$= -2\sum_{m=1}^n \epsilon_{ba} \mathcal{Q}_i^b(y') \delta(x-y'). \tag{53}$$

Therefore,

$$\partial_\mu^x \int dy' \left( Q_i^{\mu a}(x,y') + \int dy\, \Pi_a^\mu(x,y) Q_i^{ab}(y,y') \right) \Pi_b^\nu(x',y') = -2 \int dy'\, \mathcal{Q}_i^b(y') \epsilon_{ba} \Pi_a^\nu(x',y') \delta(x-y'),$$
$$= -2 \mathcal{Q}_i^b(x) \epsilon_{ba} \Pi_a^\nu(x',x). \tag{54}$$

Finally, the contraction of the second-order fluctuation is:

$$\partial_\mu^x \Xi_a^{\mu\nu}(x,y,x') = 2\epsilon_{ab} \delta(x-y) \left.\frac{\delta \Delta_b(x)}{\delta A_\nu(x')}\right|_{A\to 0} = 2\epsilon_{ab} \delta(x-y) \Pi_b^\nu(x',x). \tag{55}$$

An alternative derivation of this result will follow from the contraction of Eq. (30) after repeated application of the formula in Eq. (51). Combined with the relation $\mathcal{Q}_i^a(y) = \frac{\delta S_i}{\delta \Delta_a^{\rm mf}(y)}$, the third line of Eq. (24) is

$$\partial_\mu^x \int dy\, \frac{\delta S_i}{\delta \Delta_a^{\rm mf}(y)} \Xi_a^{\mu\nu}(x,y,x') = \int dy\, \mathcal{Q}_i^a(y) \left( \partial_\mu^x \Xi_a^{\mu\nu}(x,y,x') \right),$$
$$= 2\mathcal{Q}_i^a(x) \epsilon_{ab} \Pi_b^\nu(x',x). \tag{56}$$

This relation is opposite the one on the second line of Eq. (24), after relabeling dummy indices. Thus, we have produced the relations presented in the previous section, and the full response is gauge invariant.

---



\end{document}